\documentclass{article}
\usepackage{spconf,amsmath,graphicx,hyperref, bm, amssymb, siunitx, caption, subcaption}
\usepackage[T1]{fontenc}
\usepackage{adjustbox}

\usepackage[table]{xcolor} 
\usepackage{tabularx}
\usepackage{booktabs}

\usepackage{fancyhdr}
\fancypagestyle{firstpage}{
  \fancyhf{}

  \fancyfoot[C]{\scriptsize Jackie Lin, Jiaqi Su,  Nishit Anand, Zeyu Jin, Minje Kim, Paris Smaragdis,\\``Gencho: Room Impulse Response Generation from Reverberant Speech and Text via Diffusion Transformers,"\\In Proc. of the IEEE International Conference on Acoustics, Speech and Signal Processing (ICASSP), Barcelona, Spain, 2026.}
}
\fancypagestyle{plain}{
  \fancyhf{}

}
\newcommand{\grayhline}{\arrayrulecolor{gray!50}\hline\arrayrulecolor{black}}

\title{Gencho: Room Impulse Response Generation from Reverberant Speech and Text via Diffusion Transformers}
\name{Jackie Lin$^{1,2}$, Jiaqi Su$^{2}$,  Nishit Anand$^{2,3}$\thanks{Work done while Jackie Lin and Nishit Anand were in Adobe Internship.}, Zeyu Jin$^{2}$, Minje Kim$^{1}$, and Paris Smaragdis$^{4}$ }
\address{$^{1}$University of Illinois Urbana-Champaign, $^{2}$Adobe Research, $^{3}$University of Maryland, $^{4}$MIT}

\begin{document}
\ninept
\maketitle
\begin{abstract}
Blind room impulse response (RIR) estimation is a core task for capturing and transferring acoustic properties; yet existing methods often suffer from limited modeling capability and degraded performance under unseen conditions. Moreover, emerging generative audio applications call for more flexible impulse response generation methods. We propose Gencho, a diffusion-transformer-based model that predicts complex spectrogram RIRs from reverberant speech. A structure-aware encoder leverages isolation between early and late reflections to encode the input audio into a robust representation for conditioning, while the diffusion decoder generates diverse and perceptually realistic impulse responses from it. Gencho integrates modularly with standard speech processing pipelines for acoustic matching. Results show richer generated RIRs than non-generative baselines while maintaining strong performance in standard RIR metrics. We further demonstrate its application to text-conditioned RIR generation, highlighting Gencho’s versatility for controllable acoustic simulation and generative audio tasks.
\end{abstract}
\section{Introduction}
\label{sec:intro}

Room impulse responses (RIRs) are filters that capture the core acoustic properties of an environment, including reverberation and coloration, through a compact, parametric representation. They provide essential auditory context that shapes how sound is perceived in a given space, contributing to the realism and immersiveness of audio content. RIR estimation provides a natural basis for acoustic matching, which aims to transfer the acoustics of a reference space to new audio so that it blends seamlessly with the original scene. 

Recently, the proliferation of publicly available audio crafting tools, such as speech enhancement and text-to-speech (TTS) synthesis, has further increased the need for more accurate, flexible acoustic matching. 
This capability is critical for tasks such as automated dialogue replacement (ADR), dubbing, and voiceovers, to ensure perceptual consistency of newly recorded or synthetic speech with the original context. Likewise, TTS integrates with acoustic matching to render voices consistently in a chosen environment. Processed speech from enhancers can sound overly dry, and benefits from restoring natural room acoustics. In many scenarios, explicitly estimating RIRs is often preferred over end-to-end acoustic matching, as it produces reusable filters that can be stored, edited, shared, and applied across tasks without altering the underlying audio.

Moreover, with the emerging volume of generative tools and synthetic content, the scope of IR estimation is expanding beyond acoustic matching to a real audio reference. Applications such as immersive storytelling, AR/VR, and text-to-audio generation require \textit{soft} acoustic matching: the ability to generate diverse, semantically appropriate RIRs from weak or indirect cues—images, videos, or natural-language descriptions—in order to create coherent and immersive virtual acoustic environments. Here, generating acoustics separately from content allows users to modify acoustics without affecting speaker identity or other semantic properties and vice versa.

Despite much work, estimating RIRs from audio recordings in a blind setting—with no prior knowledge of the room, recording setup, or source signal—remains a fundamental challenge in audio processing. Lightweight parametric models are grounded in strong inductive biases, limiting their ability to capture the complexity of real RIRs. For instance, blind parameter estimation from reverberant speech~\cite{9052970, 10694126, 10094848} predicts octave-band reverberation times (T60) which can be used to drive DSP-based reverberators; however, the coarseness of these parameters limits the diversity of the synthesized RIRs. A more recent deep learning approach, Filtered Noise Shaping (FiNS)~\cite{steinmetz_filtered_2021}, constructs RIRs by upsampling a learned vector in time domain to produce an early reflection waveform and noise envelopes for the late tail. However, the output space remains constrained by the model formulation, often producing similar-sounding or unreasonable RIRs on unseen reverberant speech. In general, non-generative systems fail to capture the multi-modal nature of the blind inverse problem, and are prone to degeneracy when faced with unfamiliar inputs. As a response, GAN-based~\cite{10094770} and language modeling-based~\cite{10248189, Ratnarajah_2024_CVPR} generative models have been introduced to mitigate these issues, yet fully closing the perceptual gap between synthesized and real acoustics remains an open challenge.

We introduce Gencho\footnote{Listening examples at \href{https://linjac.github.io/Gencho/}{https://linjac.github.io/Gencho/}} (\textbf{Gen}erative \textbf{Echo}), a blind room impulse response estimator using diffusion transformers. We address the limitations of non-generative acoustic matching methods with our method's strong generalization capability and ability to produce diverse, in-distribution RIRs. Moreover, Gencho is designed to work with modern audio technologies; by leveraging speech enhancement and source separation to extract dry and early reflected speech signals, our pipeline focuses specifically on RIR estimation while integrating seamlessly into end-to-end workflows. Lastly, Gencho's generative formulation naturally supports soft acoustic matching involving different controls. Our contributions are as follows:
\begin{itemize}
    \item We propose Gencho, a complex spectrogram-based diffusion-transformer that generates diverse, plausible RIRs from reverberant speech.
    \item We propose an enhanced design of a reverberation-structure-aware audio encoder that improves the model's matching accuracy and generalization.
    \item We evaluate the designs and demonstrate Gencho's flexible applications to acoustic matching, RIR completion with hybrid-prompting, and text-to-RIR generation.
\end{itemize}

\section{Method}
\label{sec:method}

\subsection{Blind Room Impulse Response Estimation}
\begin{figure}[t!]
    \centering
    \begin{subfigure}[t]{0.65\columnwidth}
        \centering
        \includegraphics[height=1.2in]{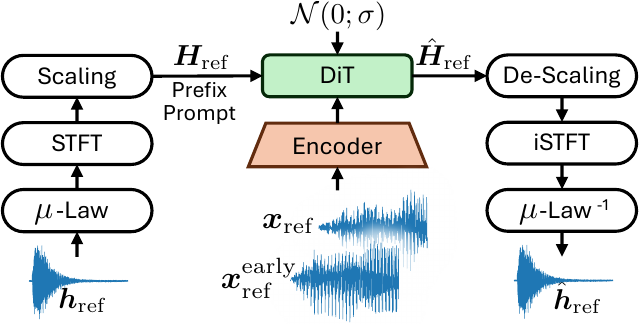}
        \caption{}
    \end{subfigure}%
    \hfill
    \begin{subfigure}[t]{0.22\columnwidth}
        \centering
        \includegraphics[height=1.2in]{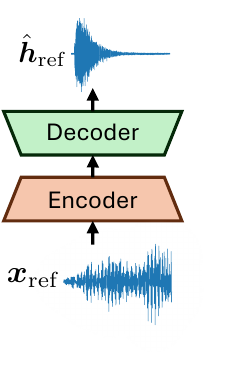}
        \caption{}
    \end{subfigure}
    \caption{(a) Model architecture of Gencho, the proposed generative estimator. (b) The non-generative FiNS-based baseline.}
    \label{fig:model}
\end{figure}

We tackle the problem of blind RIR estimation, which takes a reference reverberant speech $\bm x_\text{ref}=\bm h_\text{ref}*\bm s_\text{ref}$ and estimates the impulse response $\bm h_\text{ref}$ as a waveform signal without accessing the corresponding clean speech $\bm s_\text{ref}$ or prior information about the acoustic space. The estimated RIR is then convolved with the source content $\bm s_\text{source}$ to apply the acoustics acquired from the reference recording. 
Our blind RIR estimator aims to work on \SI{48}{kHz} audio and assumes a \SI{1.0}{s} duration for the impulse response. It consists of the structure-aware audio encoder and diffusion-based generative decoder. Figure~\ref{fig:model} illustrates a schematic diagram of our generative approach and the non-generative model variants explored in this paper.

\subsubsection{Reverberation Structure-Aware Audio Encoder}
We first encode the reference recording from the target acoustic environment $\bm x_\text{ref}$ into a 128-dimensional latent embedding $\bm w_\text{ref}$ that captures the environment's acoustic characteristics.  To this end, we adopt the time-domain convolutional encoder design from FiNS\cite{steinmetz_filtered_2021} that consists of convolutional blocks, each consisting of a 1-D convolution with a kernel size of 15 and a stride of 2, and PReLU activation. It uses adaptive average pooling at the final output of the blocks to aggregate time frames into a global latent embedding. As a modification, we replace the original batch normalization at every block with a layer normalization to handle fatal failure cases, e.g., degenerate predictions when the input audio contains a significant amount of silence. Note, as shown in Fig.~\ref{fig:model}, that the diffusion-based model's decoder is conditioned by the embedding vectors from the audio encoder also used in the non-generative method.

Our key design improvement to the audio encoder is separating the early reverberation component from the input audio. 
One challenge in blind estimation is to accurately model individual acoustic properties without mixing them up. An RIR signal can be decomposed into two parts, $\bm h_\text{ref}=\bm h_\text{ref}^\text{early}+\bm h_\text{ref}^\text{late}$, which display drastically different structures in the early reflection component $\bm h_\text{ref}^\text{early}$ (i.e., sparse) and the late tail component $\bm h_\text{ref}^\text{late}$ (i.e., diffuse noise alike). 
Different kinds of nuanced errors, while displaying reasonable distance in the waveform or spectrogram space, could lead to drastically different perceptions by listeners. For example, early reflection errors contribute to the coloration and the sense of distance, while differences in reverb time are more noticeable when the tail is short. 

Recent advances in audio separation and speech enhancement technologies enable fairly accurate isolation of foreground speech from background noise, with the flexibility to treat the perceptually sensible speech as foreground while leaving diffuse-noise-alike reverb tails to the background noise. Drawing on this capability and loosely inspired by the common practice of modeling the early reflection and the late tail of an IR separately~\cite{steinmetz_filtered_2021, 10248189}, we depart from prior methods that process the reverberant speech signal as a whole. Instead, we leverage a speech enhancement tool to extract the early-reflected component $\bm x_\text{ref}^\text{early}=\bm h_\text{ref}^\text{early}*\bm s_\text{ref}$ from the reverberant speech $\bm x_\text{ref}$. The extracted component is a relatively dry speech preserving the early reflection (e.g., the first 50ms of the reverberation) and coloration of the input. Based on this, the audio encoder takes in a two-channel input, comprising the extracted early component $\bm x_\text{ref}^\text{early}$ and the full reverberant speech ${\bm x}_\text{ref}$, and encodes it into a global embedding ${\bm w}_\text{ref}$. This structured input enables the model to explicitly analyze different components of reverberation, resulting in more accurate and perceptually realistic room impulse response estimation.

\subsubsection{Diffusion-based Generative Decoder}
Diffusion-based models have drawn attention due to their generation capabilities in domains such as images~\cite{rombach2022high} and audio~\cite{evans2024stable, novack2024ditto}.
Diffusion models employ a forward process to add Gaussian noise to the data representation and learn a backward process to remove the noise via a neural network. In this work, we introduce a diffusion-based decoder that conditions on the global embedding $\bm w_\text{ref}$ from the audio encoder and generates the complex spectrogram of impulse responses using a diffusion process. The diffusion model enables modeling of the complex distribution of impulse response signals, producing various plausible yet perceptually consistent impulse responses. This approach captures the natural variability of room acoustics, avoiding the collapse behavior observed in the conventional regression-based methods. 

\noindent\textbf{Complex Spectrogram Representation.}
Since RIR's spectrogram coefficients are sparse and of high variance, input normalization is essential: all one second-long target RIRs $\bm h_\text{ref}$ are $\mu$-law encoded in the waveform domain to emphasize low amplitudes, transformed via short-time Fourier transform (STFT) with a frame size of 128 and a hop size of 64 samples to yield a complex spectrogram ${\bm{H} \in \mathbb{C}^{65\times751}}$. Each spectrogram coefficient $c$ is then power compressed $\tilde{c}=\beta|c|^{\alpha}e^{j\angle c}$ with empirically chosen $\alpha=0.3$ and $\beta=2$. The final input dimension is $\mathbb{R}^{130\times751}$ where the real and imaginary components are separated and stacked. 
We also experimented with a latent diffusion transformer using VAE latents learned from generic audio.
However, it did not work as well as complex spectrogram diffusion, likely because the semantic relationships and distances in the general audio VAE space do not correlate strongly with perceptual distances in RIRs.

\noindent\textbf{Diffusion Formulation.}
We use the \textit{v-prediction} reparameterization~\cite{salimans2022progressive} for the diffusion training objective, which has been shown to be more stable and effective than the naive reconstruction objective in producing high-quality outputs in audio generation domains~\cite{hoogeboom2023simple}. The model structure is a Diffusion Transformer (DiT)~\cite{peebles2023scalable} that offers significant advantages in scalability and robustness. The model conditions on the timestep embedding of the current diffusion step via adaptive layer normalization, and consists of transformer layers, each comprising RMS normalization, self-attention, cross-attention, another RMS normalization and a linear layer. It cross-attends to the encoded audio embedding from the encoder to guide the generation. We use classifier-free guidance during training, dropping out the condition with 10\% probability, to encourage the model to learn both the conditional and unconditional distribution of RIRs. 

\noindent\textbf{Conditioning.} We condition the diffusion transformer on the global embedding from the audio encoder via cross-attention. During training, we experimented with two initialization setups using the pre-trained audio encoder obtained from the non-generative FiNS-like model training: training from scratch and warm initialization. Warm initialization overall shows faster convergence and more stability.

\noindent\textbf{RIR Completion and Hybrid Approach.}
During training of the diffusion model, we apply prefix prompting with a 50\% probability, replacing up to the first 50ms of diffusion latent frames with the corresponding ground-truth RIR. This encourages the model not only to learn to generate an RIR from scratch, but also to "complete" an RIR from its early reflection component (i.e., as in~\cite{lin2025deep}), offering versatile capabilities in downstream tasks. The model generates various versions of tails that sound perceptually coherent with the given partial RIR signal, allowing diverse data augmentation for acoustic simulation. We also develop a hybrid approach (see Sec.~\ref{sec:results}) that combines the benefits of the non-generative approach and the generative approach: we use the FiNS-like formulation to predict the early reflections while using the diffusion model to complete the tail.

\section{Experiments}

\subsection{Experiment Setup}
\noindent\textbf{The Baseline Model Variants.}
Three variants of FiNS were implemented as the regression-based baselines: (1) the original model, (2) a version with batch normalization layers replaced by layer normalization as described above, and (3) the layer normalization variant modified for two-channel input (the separated early reflected speech and the reverberant speech). Training was performed with a batch size of 256 for 60k steps using the AdamW optimizer with exponential decay scheduling starting at $lr=0.0001$ with decay rate $\gamma=0.999996$. We use the multiresolution STFT loss following the original FiNS.

\noindent\textbf{The Diffusion Model Variants.} The transformer-based backbone of the diffusion model has a hidden size of 256, comprised of 8 layers with 8 attention heads each, with learned positional embeddings, qk normalization, and a dropout probability of 0.1. Two diffusion model variants were implemented: (1) a single-channel model and (2) a dual-channel model. The diffusion encoders were warm-initialized with the trained encoder weights from FiNS variants (2) and (3), respectively, as they share the same encoder architecture.
The diffusion model was trained for 100k steps with batch size $=256$ on 8-gpu A100 using an AdamW optimizer and with linear cosine scheduling starting at $lr=0.0001$ and decaying to $0.00001$.

For inference, we adopt a DPM++ 2M SDE sampler with a Karras~\cite{karras2022elucidating} noise schedule, using a log-SNR range of +5.0 to –8 and $\rho=7.0$ to shape the noise decay, along with 24 sampling steps and a classifier-free guidance scale of 3.0.

\subsection{Datasets}
\noindent\textbf{Training Datasets.} The models were trained using a collection of room impulse responses from OpenSLR28~\cite{7953152}, the MIT IR Survey dataset~\cite{doi:10.1073/pnas.1612524113}, EchoThief~\cite{echothief}, Arni~\cite{prawda_karolina_2022_6985104}, dEchorate~\cite{carlo2021dechorate}, and the ACE Challenge dataset~\cite{7336912}. Since many of these datasets contain only a dozen unique rooms, but thousands of similar-sounding RIRs (mainly due to variations in microphone placement), we balanced the training data by weighting the contribution of each dataset according to its number of unique rooms, preventing overrepresentation of any single acoustic environment. Reverberant speech inputs were created by convolving these RIRs with clean speech from LibriTTS-R~\cite{koizumi23_interspeech}, a 585-hour high-quality dataset that was further upsampled to \SI{48}{kHz} using bandwidth extension~\cite{9413575}. To increase acoustic diversity, we applied data augmentations to RIRs and clean speech signals based on prior work~\cite{genhancer, su2021hifi}, randomly modifying the gains of direct-to-reverberant ratio and reverberation time durations for RIRs as well as scaling clean speech with varying speeds and volumes. All RIR signals were standardized to one second in length and normalized to a unit direct arrival energy at \SI{2.5}{ms}, and the reverberant speech inputs consist of 5-second segments.

\noindent\textbf{Evaluation Datasets.} 
We evaluated the models on an out-of-domain dataset to examine their generalization capabilities. The test set consists of RIRs from BUTReverbDB~\cite{8717722} and OpenAIR~\cite{murphy2010openair:} convolved with clean speech from the DAPS dataset~\cite{mysore2014can}, which in total span over 50 different environments, containing different signal characteristics and acoustic distributions from the training time.

\section{Results and Discussion}
\label{sec:results}
\begin{figure}[b]
    \centering
    \vspace{-4mm}
    \centerline{\includegraphics[width=1\linewidth]{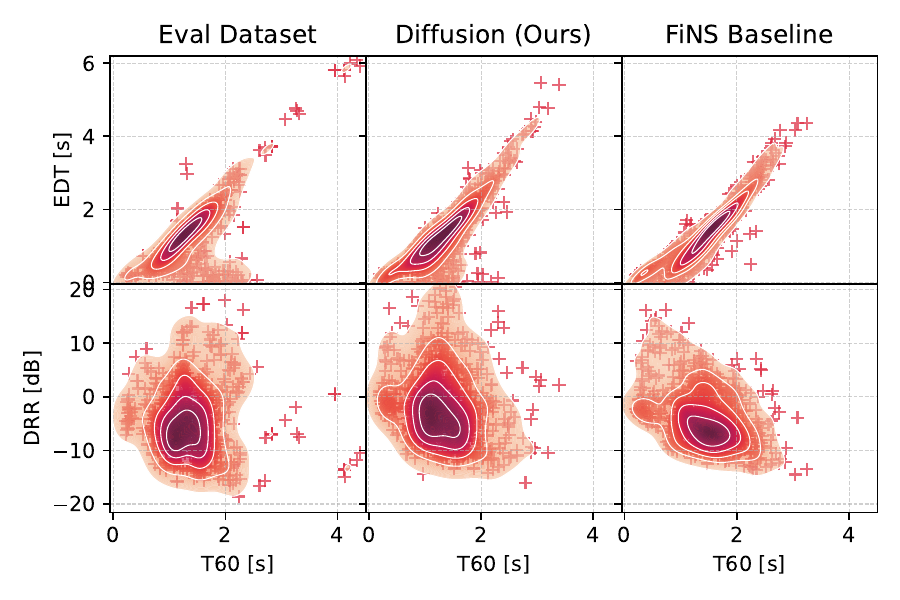}}
    \caption{Distribution of T60 vs EDT, and T60 vs DRR of the evaluation, our generated, and the FiNS layernorm generated samples.}    
    \label{fig:irstats_distrb}
\end{figure}

\begin{table*}[t]
\centering
\caption{Comparison of the non-generative baseline and our diffusion model across T60, EDT, DRR, and C50 metrics.}
\label{tab:metrics-outofdomain}
\renewcommand{\arraystretch}{1.1}
\begin{adjustbox}{max width=\textwidth}
\begin{tabular}{l|ccc|ccc|cc|cc}
\toprule
& \multicolumn{3}{c}{T60} & \multicolumn{3}{c}{EDT} & \multicolumn{2}{c}{DRR} & \multicolumn{2}{c}{C50} \\

Model & PAE (\%) & MSE (s) & MAE (s)
      & PAE (\%) & MSE (s) & MAE (s)
      & MSE (dB) & MAE (dB)
      & MSE (dB) & MAE (dB) \\
\midrule
FiNS (original)
  & 52.3 & 1.33 & 0.68
  & 45.1 & 2.27 & 0.43
  & 15.92 & 2.79
  & 53.10 & 3.47 \\

FiNS layernorm
  & 22.6 & 0.19 & 0.31
  & 20.1 & 0.14 & 0.22
  & 8.93 & 2.38
  & \textbf{3.69} & 1.45 \\

FiNS layernorm + 2ch
  & 14.2 & 0.10 & 0.20
  & \textbf{15.5} & \textbf{0.08} & \textbf{0.16}
  & \textbf{8.86} & \textbf{2.29}
  & 3.89 & \textbf{1.41} \\

\grayhline

Gencho + 1ch
  & 16.3 & 0.13 & 0.23
  & 26.4 & 0.22 & 0.28
  & 25.91 & 4.11
  & 12.15 & 2.69 \\

Gencho + 2ch
  & 13.6 & \textbf{0.08} & \textbf{0.18}
  & 34.8 & 0.16 & 0.25
  & 25.62 & 4.04
  & 11.81 & 2.66 \\

\midrule \midrule
Hybrid + 2ch @5ms
  & \textbf{13.5} & 0.085 & 0.184
  & 23.6 & 0.14 & 0.23
  & 26.75 & 3.53
  & 14.16 & 2.63 \\

Hybrid + 2ch @25ms
  & 13.8 & 0.085 & 0.185
  & 25.5 & 0.13 & 0.23
  & 23.78 & 3.40
  & 12.09 & 2.45 \\
\bottomrule
\end{tabular}
\end{adjustbox}
\vspace{-3mm}
\end{table*}

We report the standard RIR metrics of the methods. For each target-generated RIR pair, we compute the reverberation time to -60dB (T60), early decay time to -10dB (EDT), direct to reverberant ratio (DRR), and clarity index (C50). The percentage absolute error (PAE) of each time-related metric, and mean squared error (MSE) and mean absolute error (MAE) of every metric are reported in Tab.~\ref{tab:metrics-outofdomain} for the out-of-domain test set. First, we see that the original FiNS does not perform well on new, unseen data; the layernorm and the structure-aware encoder improve performance, achieving 15.5\% PAE on the EDT, \SI{2.29}{dB} and \SI{1.41}{dB}. Our diffusion-based model Gencho+2ch (taking in both the early reflected speech and the full reverberant speech) shows benefits in more accurately modeling the reverberation time, achieving the lowest T60 error with 13.6\% PAE and 0.18s MAE, but degrades on EDT, DRR, and C50. 
We hypothesize that diffusion models are effective at modeling the randomness of noise-like tails in reverberation, but their inherent stochasticity may introduce excessive variability for early reflections, which tend to have clearer and more structured patterns.

Meanwhile, the RIR statistics on the evaluation dataset plotted in Fig.~\ref{fig:irstats_distrb} show that our generative model is more expressive and approximates the underlying distribution of the target RIRs more closely than the non-generative baselines. The joint distribution of T60 and DRR reveals that FiNS exhibits a stronger inverse correlation between the two. This is likely a result of "regression to the mean", where the model tends to default to the common, averaged pattern that less reverberant environments are typically smaller and therefore associated with higher DRRs.


\noindent\textbf{Hybrid approach.} As observed, the non-generative baseline achieves more accurate DRR, whereas the diffusion-based approach produces more realistic decaying reverberation tails as reflected by improved T60 accuracy. Motivated by these complementary strengths, we construct a hybrid approach leveraging the IR completion capability of the trained diffusion model. We first estimate the RIR using the improved FiNS, then feed the early portion of that as a prompt to Gencho to generate the remaining RIR. We report the performance of this hybrid approach for prompt lengths~=~\SI{5}{ms} and \SI{25}{ms} in Tab.~\ref{tab:metrics-outofdomain}. The results show that this hybrid approach improves the accuracy of the generative model in EDT and DRR while maintaining T60 accuracy.

\section{Text-to-RIR Generation}

Our diffusion-based RIR generator enables the creation of diverse, novel impulse responses under weak guidance and can be extended to a broad range of multi-modal applications. 
It offers flexible controls over acoustic properties and allows users greater freedom to explore virtual acoustic spaces. 
To demonstrate, we adapt the model for text-conditioned RIR generation (i.e., text-to-RIR) by replacing the audio encoder with a Flan-T5-XXL\footnote{https://huggingface.co/google/flan-t5-xxl} text encoder. The diffusion model cross-attends to the text embedding sequence. This allows natural-language descriptions of an acoustic environment, e.g. “a large cathedral with long echoes and a distant human talker", to condition the diffusion process to generate semantically matching RIRs. 

The training dataset comprises approximately 150K pairs of impulse response signals and their corresponding text captions generated by ParaLLM~\cite{parallm} using a similar set of RIRs as in the aforementioned experiments. Each caption describes the semantic interpretation of the acoustic characteristics of the RIR in natural language, such as the perceived spaciousness of the room, the speaker's apparent distance, and the clarity of the speech. We fine-tune a pre-trained Gencho decoder together with the newly introduced text encoder to enable transfer learning and accelerate convergence. We then evaluate the model on the validation RIR-caption pairs held-out from training as well as on a set of newly generated RIR-caption pairs covering unseen RIRs from OpenAIR~\cite{murphy2010openair:}. 

\begin{figure}[htb]
    \centering
    \vspace{-2mm}
    \includegraphics[width=\linewidth]{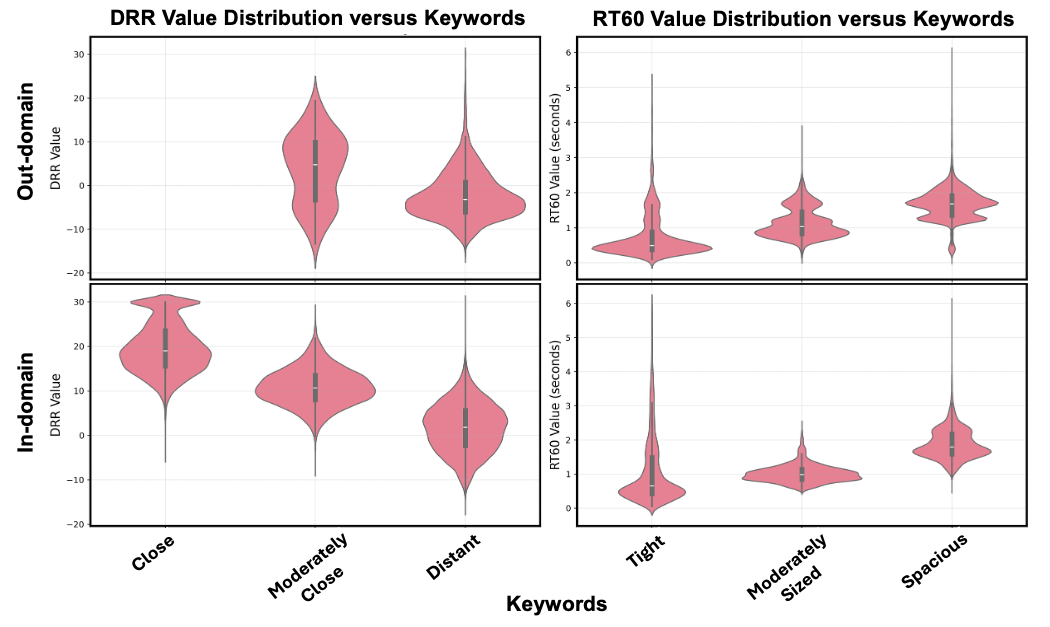}
    \vspace{-5mm}
    \caption{Violin plot. x-axis is short text prompts perceptually related to reverberance. y-axis is the DRR and T60 of generated RIRs.}
    \label{fig:text2ir}
    \vspace{-5mm}
\end{figure}

Fig.~\ref{fig:text2ir} illustrates the strong correlations between reverberance-related keywords in the text descriptions and the acoustic properties (DRR and T60) of the generated RIRs across both test sets. Each keyword results in a differently centered distribution of acoustic properties that is semantically consistent with the keyword. For example, the keyword "tight (space)" leads to shorter reverberation time, while "spacious" leads to longer reverb tails. OpenAIR does not contain closely captured impulse responses, so the keyword "close" does not appear in its captions. 
We further evaluate the accuracy based on the acoustic property bins designed in ParaLLM: \textbf{DRR Level} as DRR values divided into $\leq 5$, $5 \sim 11$, and $\geq 11$ db ranges, \textbf{T60 Level} as $\leq 0.5$, $0.5 \sim 1.2$, and $\geq 1.2$ in seconds, and lastly a 3-by-3 grid of overall \textbf{Reverb Level} crossing DRR level and T60 level. We generated five variations of RIRs per text caption, achieving average accuracies of $70.1\%$ and $89.1\%$ on \textbf{DRR Level} for in-domain and out-domain test sets, $82.7\%$ and $85.1\%$ on \textbf{T60 Level}, and $57.2\%$ and $76.3\%$ on \textbf{Reverb Level}. The results show strong alignment with the ground-truth categories of the RIRs paired with the text captions, while the in-domain test set generally yields lower accuracies due to its broader thus more challenging coverage of acoustic environments.
We refer readers to our demo page to check out the text-to-RIR examples. 

\section{Conclusion}
We introduced Gencho, a diffusion transformer for blind room impulse response estimation that generates complex spectrogram RIRs from reverberant speech. The model leverages a structured input of separated early and late reflections to improve the robustness of conditioning information, and generates diverse, perceptually realistic outputs via a diffusion process. Gencho integrates seamlessly with standard speech-processing pipelines and can be adopted to a range of novel controllable tasks, including RIR completion and text-conditioned RIR generation. Experiments demonstrate improved generalization and richer acoustic statistics compared to non-generative baselines, establishing Gencho as a flexible tool for controllable acoustic simulation and generative audio applications.

\bibliographystyle{IEEEbib}
\bibliography{references}

\end{document}